\documentclass[letterpaper, conference]{IEEEtran}
\IEEEoverridecommandlockouts
\usepackage{cite}
\usepackage{amsmath,amssymb,amsfonts}
\usepackage{algorithmic}
\usepackage{graphicx}
\usepackage{textcomp}
\usepackage{xcolor}
\def\BibTeX{{\rm B\kern-.05em{\sc i\kern-.025em b}\kern-.08em
    T\kern-.1667em\lower.7ex\hbox{E}\kern-.125emX}}
\begin{document}

\title{A Comprehensive Survey on Real-Time Voltage Stability Assessment for Power Systems}

\author{\IEEEauthorblockN{Gourav Wadhwa \hspace{3pt} Amandeep Kharb \hspace{3pt} Satyam Mishra \hspace{3pt} Mohit Kumar \hspace{3pt} Shreyansh Srivastav}
\IEEEauthorblockA{\textit{Department of Electrical Engineering} \\
\textit{Indian Institute of Technology, Ropar}\\
\{2017eeb1206, 2017eeb1128, 2017eeb1165, 2017eeb1152, 2017ceb1028\}@iitrpr.ac.in}}


\maketitle

\begin{abstract}
Accurate real-time assessment of power systems voltage stability has been an active area of research in the past few decades. In the past decade, after the development of phasor measurement units (PMU), a lot of discussions has been going on phasor measurement techniques for real-time voltage stability. The fundamental idea behind these methods is to find the Thevenin equivalents of the system, and then determine the voltage stability margin based on the equivalent circuits. Some approaches also include the use of Artificial Neural Networks (ANN), for online monitoring of voltage stability margins. These methods are really fast as compared to the other methods. It has been shown that if we can obtain the phase angles and voltage magnitude in real-time from the phasor measurement units (PMU), then the voltage stability margins can be obtained in real-time and we can initiate voltage stability control methods. We are going to discuss Thevenin's equivalent methods and Artificial Intelligence methods in detail in this paper. We will also introduce the traditional methods which were earlier used for power systems stability assessment such as Time Domain methods, Static Methods, and Sensitivity methods. We are going to finally compare these methods and try to give general guidance on choosing a power stability method.
\end{abstract}

\begin{IEEEkeywords}
Power Systems, Voltage Stability, Artificial Neural Networks, Thevenin's Equivalent
\end{IEEEkeywords}

\section{Introduction}
With the increase of competition between the power producing companies, electric power grids are mostly operated near the security limits of the power system, because of which the risk of instabilities such as voltage instability, angle instability, and frequency stability \cite{ref-1, ref-2} has significantly increased in the past few decades. Voltage stability is one of the major concerns as it may propagate in the system and produce cascading effects, due to which it may affect the whole system. With these risks, the planning of voltage instabilities cannot be made few hours ahead of the systems because of which it is important to find an online real-time voltage stability assessment method that could determine the voltage stability margins.

\subsection{Classification}

Voltage Stability also known as load stability is a major concern for power systems as it has potential to damage the system by producing cascading effects. \cite{ref-3}. A lot of methods have been proposed to trace the voltage stability margins. These methods can be separated into different classes on the basis of the approach to the problem :
\subsubsection{Artificial Neural Networks (ANN)}
Artificial Neural Networks (ANN) are used to find non-linear functions that could map the raw input data to meaningful output data. These approaches are fast and very accurate to predict voltage stability margins. In \cite{ref-4}, proposed an ANN-based method for quickly finding the voltage stability margins for the power systems. The real time voltage stability margins were found out by accessing the real time voltage magnitudes and phase angles through phasor measurement units (PMU). We will further discuss this approach in great detail in this paper.
\subsubsection{Thevenin Equivalent}
These methods are also very fast, and there has been a lot of discussion regarding these methods \cite{ref-5, ref-6, ref-7, ref-8}. These methods use wide-area measurement or local area measurement to identify the thevenin equivalent of the power system. We will discuss state-of-the-art approaches that uses this method.
\subsubsection{Energy Functions}
In \cite{ref-9}, the authors use Lyapunov Energy Function to find the voltage stability margins by the variable gradient method. This approach was based on the Synchronous machine infinite bus system (SMIB), for which they used non-linear Lyapunov Energy function. This method works really good for small power systems; however, it has been shown that these method fails to accurately predict the voltage stability margins for realistic size power systems.   
\subsubsection{Static Methods}
These are the traditional methods that have been tested for a long time. These methods can successfully determine the voltage stability limit using many different techniques such as the singular value decomposition method, continuous power flow method, sensitivity analysis method, and collapse point method with the summary and comment \cite{ref-10}. These methods have shown a lot of progress in the past decade.
\subsubsection{Time-Domain Methods}
In \cite{ref-11}, Shahooei et al. use an EMTP-based time-based approach to simulate the voltage stability of the system. They used time changing resistive, inductive, and capacitive loads to model the power system. Then they plotted the PV curves for these loads and try to predict the voltage collapses in the system. As we can notice from the above example, these models need a high degree of detail about the model, and these methods are slow compared to other methods, hence these methods are not used in the modern power system for predicting voltage collapses \cite{ref-12}. 

\subsection{Background}
There has been a lot of discussion about all the above-mentioned methods to determine the voltage stability margin. In this paper, we will only discuss two types of methods, thevenin equivalent method, and the artificial neural network voltage stability margin determination method. In this section, we will briefly discuss the basics of these two methods.

Artificial Neural Networks (ANN) are used to determine the non-linear function that can map multi-dimensional input to multi-dimensional output. These methods are currently being used to solve many difficult problems such as air-traffic control problems, computer vision tasks, automatic driving, etc\cite{ref-13, ref-14, ref-15}. Most of these methods learn by observing the examples of input repeatedly and tries to guess the correct answer for these inputs. These methods can be very accurate and very fast if we provide the correct input to these networks. In \cite{ref-4}, the authors find the voltage magnitude, voltage angles, real power, and reactive power for all the nodes using the power flow equations, and then input these values in the neural network which can capture the non-linearity in the system. To train the neural network they gathered a sufficiently large number of load/generation levels randomly (Conventional Flow programs have been used to gather these cases in order to mitigate any discrepancy in generated data and real-world data).

Another very popular method for real-time voltage stability assessment is thevenin's equivalent method. The most critical and crucial part of this method is finding a proper thevenin equivalent of the power system.Figure-1 shows a 2-bus system. In Figure-1, we consider a 2-bus power system, in this system we consider that the load bus k is connected to a load which consumes $\Bar{S}_k = P_L + j Q_L$ power where $P_L$ is the active power consumed by the load and $Q_L$ reactive power consumed by the load. Let the voltage at the load bus k be $\Bar{V}_k$ and the voltage at the generator end be $\Bar{V}_{thev}$. The generator is connected to the load bus k through an impedance of $\Bar{Z}_{th}$. We can clearly see that the current $\Bar{I}_k^*$ can be represented as follows:
\begin{equation}
    \Bar{I}_k^* = \frac{\Bar{S}_k}{\Bar{V}_k} = \frac{(\Bar{E}_{th} - \Bar{V}_k)^*}{\Bar{Z}_{th}^*} 
    \label{eq-1}
\end{equation}
If we do some little manipulations in the above equations then we get :
\begin{equation}
    V_k^2 - \Bar{E}_{th}^*\cdot\Bar{V}_k + \Bar{S}_k\cdot\Bar{Z}_{th}^*
\end{equation}
From the above equations, we can observe that $V_k$ will have voltages solutions for this equation, either both of them will be real or both of them will be complex. When the loads are lower for the systems then we will have two solutions for the system but as we gradually increase the loads and reach the maximum transfer then we will have only one solution for this equation, beyond which we cannot transfer any power to the loads. This is $P_{max}$ for this system. This limit will be reached when:
\begin{equation}
    \begin{aligned}
        V_k &= \Bar{E}_{th}^2 / 2 \\
        \Bar{Z}_{th}^*& = \frac {V^2_k}{\Bar{S}_k} = \Bar{Z}_k^* 
    \end{aligned}
    \label{eq-3}
\end{equation}
    
From equation \eqref{eq-3} we can easily observe that the maximum power transfer is happening when the thevenin impedance matches the load impedance. Beyond this limit, we cannot transfer more power as it may damage the power system and hence this the $P_{max}$ for the system. Any further rise in demand from the load side will lead to voltage collapse and the system will become unstable. However, practical power systems are typical of very large sizes (containing more than 100 buses). To use such stability assessment techniques on multiple bus systems, one needs to analyze each bus one by one by converting the rest of the system to its thevenin equivalent model as shown in Figure-1. The determination of thevenin equivalent parameters involves several numerical techniques that are based on real-time data obtained from PMUs which are very popular in the modern power system and hence are installed at every bus of the power system. We will discuss a few methods to determine the thevenin equivalent in this paper.
\begin{figure}
    \centering
    \includegraphics[scale=0.34]{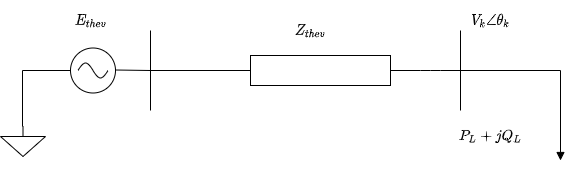}
    \caption{A simple 2-bus system where the load (consuming $P_L + j Q_L$ power)is connected to a generator with voltage $E_{thev}$ through a impedance $Z_{thev}$.}
\end{figure}

The rest of the paper is organized as follows: In section II, we discuss the methodologies using Artificial Neural Networks (ANN) for online assessment of power stability margins. In section III, we move on to further discuss different methodologies using thevenin's equivalent method for assessment of voltage stability. In section IV, we finally conclude our work for this paper. 

\section{Artificial Neural Network (ANN) methodologies}
Load P margins can be calculated directly from the PV curve; if the real power delivered at the operating value is $P_{0}$ and the maximum possible power that can be delivered is $P_{max}$ then $P_{margin}$ can be calculated as:

\begin{equation}
    P_{margin} = P_{max} - P_0
\end{equation}

\begin{figure}
    \centering
    \includegraphics[scale=0.70]{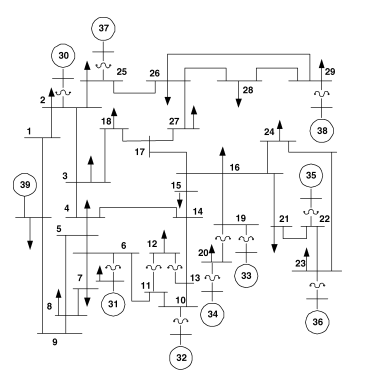}
    \caption{New-England 39-bus system}
\end{figure}

The loads are estimated using load forecasting program and according we try to change the power outputs of our generators to meet the power requirements but still there is some differences between the generation side and the loading side. In \cite{ref-4}, Zhou et al., these differences between the generation and loading side are modelled as:
\begin{equation}
    \begin{aligned}
        P_g &= P_{g0} + \lambda P_{gd} \\
        P_l &= P_{l0} + \lambda P_{ld} \\
        Q_l &= Q_{l0} + \lambda Q_{ld}
    \end{aligned}
\end{equation}

where $P_l$ is the real power delivered to the loads, $P_g$ is the real power generated by the generators, and $Q_l$ is the reactive power delivered to the loads. $P_{g0}, P_{l0},$ and $Q_{l0}$ represents the initial guess and then these are changes in the direction in which $P_{gd}, P_{ld},$ and $Q_{ld}$ and changed respectively. These changes are multiplied with a factor $\lambda$, which determines how much change will be enforced every update. The authors used the load forecast values to find the direction of $P_{ld}$, and as the reactive power components are not available in the forecast, they used a constant power factor from which they found the direction and magnitude of $Q_{ld}$. 

They used conventional power flow equation and solved them iteratively using the Gauss-Siedel method or Newton-Raphson method \cite{ref-18} to determine the voltage angle $\theta_{0}$, voltage magnitude $V_{0}$, reactive power $Q_{0}$ and active power $P_{0}$ at the initial operating point for all nodes in the power system. Using continuation power flow, we can trace the steady-state voltage stability limit, which is further used to determine $P_{margin}$ \cite{ref-19, ref-20}, but these methods are very slow and hence cannot be used in large-scale power systems. Therefore the authors use very fast methods known as Artificial Neural Networks predict to $P_{margin}$ using the initial guess and load forecast. Using ANN for calculating $P_{margin}$, we can circumvent all the calculations needed in the continuous power flow equations.


The ground truth ($P_{margin}$) for these randomly generated initial operating points used for training the neural network is done through the Continuation Power flow. Before generating these ground truths, we verify that the initial operating points are even feasible in the real world. The authors determined that using any pair of the four initial conditions for an N-bus system would be enough as the other two would be just a linear function of the chosen two. Then they developed a multi-layer perceptron (MLP) with an input layer, a hidden layer, and an output layer. The input layer contained any two variables of the four, and one hidden layer was used for this task with five neurons. They trained the network with Mean squared error loss, which is defined as follows:
\begin{equation}
    Loss = (P_{m, pred} - P_{m, orig})^2
\end{equation}
where $P_{m, pred}$ is the predicted value from the Artificial Neural Network (ANN) and $P_{m, orig}$ is the calculated value from the continuous power flow (CPF) equation. The authors reported that the loss for the network decreased below $10^{-5}$ after training for 100 epochs on the training data of size 3000 initial conditions. They used the momentum optimizer with a learning rate of 0.01 with the momentum constant of 0.9. To generate the training data they added random disturbances to the active and reactive power of the loads, and they also added rather small disturbances to the generator voltage magnitudes (The variation was considered between $\pm 30\%$ in the system).

\begin{figure}
    \centering
    \includegraphics[scale=0.70]{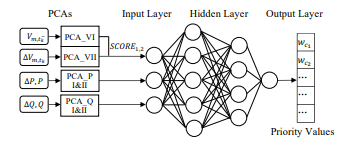}
    \caption{In \cite{ref-21}, Cai et al. used the above architecture to find the control action for the given system dynamics.}
\end{figure}

The authors tested the algorithm on a 39-bus system, as shown in Figure-3. When they calculated the error between the predicted and the ground truth values, they found that the maximum error in $P_{margin}$ was 0.6566\%, and the mean error was 0.1205\%. As the calculated values of ground truth will also be having error despite which also maximum error observed in the system with original values was observed to 1.1222\%, whereas the mean error was as low as 0.2559\%. While testing the algorithm, they observed that for the 39-bus system, it just took 0.0782 seconds to predict the $P_{margin}$, and for the Alberta 1844-bus system, it took 0.3810 seconds, which shows the efficacy of the method. As discussed previously, this method could be given any two parameters out of the four parameters according to the results from the paper they observed that minimum error was found in giving Voltage magnitudes and voltage angles as the input to the network.

In \cite{ref-21}, Cai et al. proposed a data-driven learning approach and control method to handle the voltage stability in real-time for the power system. The authors thought that different kinds of fault in the system would lead to a similar kind of control as these faults might have a similar effect on the power system. Therefore they concluded that they directly try to find the relationship between the system dynamics, which is the node voltage magnitude, voltage phase, active power, and reactive power. They used a neural network, which has shown very powerful results to map the input data to the control action. These control actions are all associated with the priority value, which can be determined from the control effects. Principal component analysis (PCA) was used to reduce the dimension of the input data. In Figure-2, we show the basic architecture of the neural network used by this paper to find the control action for the given system dynamics.

They divided their approach into three basic steps; the first one is off-line long-term optimization in which they find the optimal control actions by the offline long-term search for almost all the anticipated faults. To find the optimal control action, they used the greedy algorithm \cite{ref-22} in which they choose to solve the problem by choosing the locally optimal choice at each stage. The second step is off-line learning, in which they apply the PCA on the system dynamics and then find the control actions using artificial neural networks (ANN). While training the neural network, the loss function used is cross-entropy loss or logarithm loss. The final step is online control and learning in which they find a feasible solution from all the data they are having from the previous two methods. Using this method, we can also target the unanticipated faults that may occur due to different reasons, but the power system’s response is similar. This can be really helpful in the real-world power system. 

They also evaluated the approach on the New-England 39-bus system, as shown in Figure-7. They observed that the training accuracy for the 39-bus system was 96.80\% for the control action, and this method was able to predict a control action in 0.012 seconds. We cannot compare the previous two approaches directly because both of the approaches are finally predicting different things.

\section {Thevenin Equivalent methodologies}
In this section, we consider some very basic methods for calculating the thevenin equivalent for a given power system. These are online real-time power system assessment methods and for these methods to work we would need the measurements from PMU.

In \cite{ref-5}, Vu et al. use a very basic idea of least squares. In this, they rewrite the equation \eqref{eq-1} as:
\begin{equation}
    \Bar{E}_{th} = \Bar{V}_k + \Bar{Z}_{th}\cdot\Bar{I}_k
    \label{eq-7}
\end{equation}
by replacing thevenin voltage, $\Bar{E}_{th}=E_r+jE_i$, Bus voltage, $\Bar{V}_k=u+jw$, and Load current, $\Bar{I}_k=g+jh$. we can represent the equation-\eqref{eq-7} as matrix multiplication operation which is represent as follows:
\begin{gather}
    \begin{pmatrix}
      1 & 0 & -g & h \\ 0 & 1 & -h & -g
    \end{pmatrix}
    *
    \begin{pmatrix}
      E_r \\ E_i \\ R_{th} \\ X_{th}
    \end{pmatrix}
    =
    \begin{pmatrix}
      u \\ w
    \end{pmatrix}
\end{gather}

In the above equation, g, h,u, and w are known variables obtained from real-time measurement whereas $E_r$, $E_i$, $R_{th}$, and $X_{th}$ are unknown which represent the thevenin equivalent parameters. Since, we have two equations and four variables, therefore estimation is not possible by just one set of values of V and I. Hence we sample at multiple values of V and I at different time frames and assume that thevenin equivalent parameters do not vary until their value is obtained. Although, the authors could have obtained thevenin equivalent parameters by sampling at only two-time instants but the authors preferred to choose multiple time instants so as to account for the dynamics involved in the system. The multiple thevenin equivalent values thus obtained are fitted along the smooth curve through the Least square method which gives its variation with time.

In \cite{ref-6}, Smon et al. propose to use different forms of Tellegen's Theorem. Basically, Tellegen’s theorem states that for a given electrical network that follows KCL and KVL, the summation of electrical power in each branch of the network is zero at any instant of time. This allows us to write:
\begin{equation}
    \hat{I}^T \Delta U - \hat{U}^T \Delta I = 0
    \label{eq-9}
\end{equation}
where $\Delta U$ and $\Delta I$ represents the changes in voltage and current phasors of all buses and branches of the incremented network and $\hat{U}$ and $\hat{I}$ are voltage and current phasors of the ad-joint network. As the adjoint and incremental network have similar topology so equation \eqref{eq-9} holds perfectly. After further simplification by using equation \eqref{eq-9} as our base equation, we arrive on  the result that the maximum transfer limit of a network is reached when
\begin{equation}
    \begin{aligned}
        \hat{Z}_{th} &= \hat{Z}_k \\
        |\frac{\Delta U_k}{\Delta I_k}&| = |\frac{U_k}{I_k}| 
    \end{aligned}
\end{equation}

Using the above equations the authors defined a simple normalized impedance-stability index which can be used for comparisons with other methods. From the above equation, we can see that stability analysis can be easily done by obtaining incremental values through small disturbances to obtain $Z_{th}$. An update on the thevenin impedance is only made when  $\Delta I$ is greater than a chosen threshold value. This eliminates the case of $\Delta I = 0$. 

Carsi et at. \cite{ref-7}, proposed adaptive method to obtain the real-time online assessment of power system stability. The authors made an assumption of neglecting the thevenin resistance, $R_{th} \approx 0$. This assumption is made because at high voltage level busses we know that $X_{th} > R_{th}$. They further extended the work done by least square methods and rewrote the equation-\eqref{eq-7} as:

\begin{equation}
    \begin{aligned}
        E_{th} cos \beta &= V_k cos \theta \\
        E_{th} sin \beta = &X_{th} I_k + V_k sin \theta
    \end{aligned}
\end{equation}
However, the no. of unknowns are 3 while equations are 2 only. Hence to solve such a system we first need a rough estimation of Eth to start the first iteration. This we take as $E_{th}^0 = (E_{th}^{max} + E_{th}^{min}) / 2$. Then we put the estimated value of $E_{th}$ into the system of equations and obtain the value of remaining unknowns. As the load impedance of the bus changes, the estimated value of $E_{th}$ is updated in accordance with the given set of rules as mentioned below: \\
\\
Step 1: Estimate the $E_{th}^0$ using the average of maximum and minimum $E_th$ as given above \\
Step 2: Then calculate $X_{th}^0$ and $\beta^0$ \\
Step 3: Calculate $E_th$ according to the following rules - \\
if $dZ_k^i\cdot X_{th}^i < 0$ then $E_{th}^i = E_{th}^{i-1} + \epsilon_E$ \\
if $dZ_k^i\cdot X_{th}^i > 0$ then $E_{th}^i = E_{th}^{i-1} - \epsilon_E$ \\
if $dZ_k^i\cdot X_{th}^i < 0$ then $E_{th}^i = E_{th}^{i-1}$ \\
Step 4: Then we further calculate the values of $X_{th}^i$ and $\beta^i$ \\
Step 5: Increase the value of i and repeat the step 3 \\ \\
After obtaining the updated value we insert it again in the equation to obtain the next set of values of $X_th$. After multiple iterations, we finally obtain $X_{th}$ with greater accuracy.

In \cite{ref-8}, the authors introduce a coupled single-port circuit based method. In this method, the system is demonstrated as a multi-port coordinate with stacked buses as well as generators brought outside the system. In the network buses with no current whatsoever are taken inside. Following is the representation of the system in the matrix form:
\begin{gather}
    \begin{pmatrix}
      -I_L \\ 0 \\ I_G
    \end{pmatrix}
    =
    \begin{pmatrix}
      Y_{LL} & Y_{LT} & Y_{LG} \\ Y_{TL} & Y_{TT} & Y_{TG} \\ Y_{GL} & Y_{GT} & Y_{GG}
    \end{pmatrix}
    \begin{pmatrix}
      V_L \\ V_T \\ V_G
    \end{pmatrix}
\end{gather}

Here V, as well as I, are the symbols for vectors of phasors of voltage and current, the matrix Y represents system admittance matrix and Load bus is denoted by subscript L , tie bus by T and generator bus by G. If we eliminate the voltage vectors of the tie busses by the following equations:
\begin{equation}
    \begin{aligned}
        V_L =& K V_G - Z_{LL} I_L \\
        Z_{LL} = (&Y_{LL} - Y_{LT}Y_{TT}^{-1}Y_{TL})^{-1} \\
        K = Z_{LL} &(Y_{LT} Y_{TT}^{-1}Y_{TG} - Y_{LG})
    \end{aligned}
\end{equation}

we obtain the following equations for the load bus k:
\begin{equation}
    \begin{aligned}
        V_{Lk} &= E_{eqk} - (Z_{eqk} + Z_{couple-k}) I_{Lk} \\
        E_{eqk} = &Z_{LL_{kk}}, Z_{couple-k} = \sum_{i=1, j=k}^n Z_{LL_{ik}} \frac{I_{L_{i}}}{I_{L_k}}
    \end{aligned}
    \label{eq-14}
\end{equation}

They modeled the coupling effect of the other loads into the $Z_{couple-k}$ and found the thevenin equivalent using the equation \eqref{eq-14}. The matrix K helps us to determine the generator terminal voltage and further, we determine the thevenin impedance by using the diagonal element of $Z_{LL}$ and the coupling effects of the loads. In this study all the loads were assumed to increase at the same rate, this assumption made the system linear, and the ratio of the two buses nearly constant.

\subsection{Time Complexity}
In this section, we will discuss the time complexities of the above-presented methods for real-time online assessment of power system stability. All these methods are good candidates for real-time voltage instability assessment as they are computationally less expensive than traditional methods. So for comparison, we will analyze these methods in terms of their time complexity and amount of measurements needed. Without any loss of generality, we can consider our system is containing buses which consist of $N_L$ load buses,$N_T$ tie buses and $N_G$ generator buses respectively. For $M$ load buses, the time complexities and measurement needed to calculate thevenin equivalent are considered at every time step. 

\subsubsection{Least squares method}
In order to calculate the thevenin's equivalent of M load buses, all of the load buses should have PMUs. The time for calculating thevenin's equivalent of one bus is closely determined by the size of time window W. The time window is generally in the order of a few seconds. So we can say that for one load bus run time would be $\mathcal{O}$(W). Therefore, the time complexity for calculating M load buses are $\mathcal{O}$(MW).
\subsubsection{Tellegen's Theorem Based method}
The tellegen's theorem can evaluate the system's voltage system by just using the current and voltage phasor measurements, because of which it is very useful for PMU based online monitoring of the system. The PMU needs to be installed at M load buses and measurements can be used to calculate Thevenin Impedance using the equation as mentioned in the previous section. The run time of one load bus is O(1) and so for M load buses,it will be $\mathcal{O}$(M).
\subsubsection{Adaptive Method}
The adaptive method can recognize raising of the voltage instability phenomena in a very fast way, seen by an EHV bus provided PMUs for a real-time power system. This method has been previously determined to be satisfactory in terms of robustness, speed, and reliability for a large power system with ZIP load characteristics. To find the thevenin's equivalent this method also requires M PMUs to be installed at load buses. In this method, we need to calculate $E_{th}$ at every time step, however, given the conditional statements in the previous section we can easily run this algorithm in $\mathcal{O}$(1) for one load bus, and therefore the time complexity for M load buses will be $\mathcal{O}$(M). 
\subsubsection{Coupled-Single port circuit method}
Applying the basic impedance match technique to the multi-load system for predicting the voltage stability margin causes a few problems. Therefore to remove this problem we can use a coupled single-port circuit. This method can speed up the continuation power flow algorithms and can be used to identify critical buses without using nodal analysis. For the coupled port method, $N_G$ PMUs are needed to be installed on all generation buses in addition to  M PMUs which are used to monitor load buses. Therefore, the total number of PMUs is M+$N_G$.The run time for this method for a single bus is $\mathcal{O}$($N_G$), after which it takes $\mathcal{O}$(1) time to calculate the impedance of the system. So the overall time complexity of this part for a single load bus is $\mathcal{O}$($N_G$) which can be further extended to be $\mathcal{O}$(M$N_G$) for an M load bus.

Coupled Single port method requires extra PMUs which are needed to be installed at generator buses as compared to Least Squares, Tellegen Theorem, and Adaptive Method where PMUs are installed only at load buses. The time complexity for tellegen's theorem method, and adaptive method are estimated as $\mathcal{O}$(M) for at one-time step for M load buses. The time complexity for the least-squares and coupled single-port methods are dependent on the specific system. The major drawback for the coupled single-port method is the $\mathcal{O}$($N^3$) time complexity for large power systems. 

\subsection{Qualitative Analysis of Thevenin Methods}
\subsubsection{Least Squares}
The least-square method can be used to track the vicinity of a steady-state voltage instability by approximating the thevenin equivalent of the network but it can not find the exact point of instability as the evaluated parameters always lack their true value. As this method uses an approximation that thevenin equivalent parameters remain constant for a fixed time window of their evaluation, which for a practical non-linear and dynamic system might not be true, so to suppress such inaccuracy a larger time window is required to collect a larger number of sample measurements, which at the end increases the computational burden of the algorithm.

\subsubsection{Tellegen’s Theorem method}
Tellegen's theorem method uses a different technique than other methods where they estimate the thevenin's equivalent parameters by fitting the curve for two consecutive phasor measurements. This method is computationally very fast because the thevenin's equivalent parameters are directly identified by finding the current and voltage increments in the base network which is subjected to some disturbances. Therefore, it is also very easy to implement in a wide-area monitoring and control center. We can also use this method on fast dynamic systems to suppress oscillations of the estimated parameters. 

\subsubsection{Adaptive Methods}
The effectiveness of this algorithm lies in its ability to work at very fast sampling rates (of about 20ms) which allows it to recognize voltage instability phenomenon precisely at EHV buses provided with PMUs.

\subsubsection{Single-port coupled circuit}
This new circuit model mitigates errors that occur when a multi-load bus system (nonlinear and dynamic) is decomposed into the thevenin equivalent model. It takes into account the coupling effects of the load and deals with it explicitly by bringing it outside the equivalent system. Hence the accuracy of this model in predicting voltage instability is comparatively very high. However, the technique requires an extra number of PMUs to be installed at generator buses which increases the hardware complexity and run time of the algorithm. 

\section {Conclusion}
In this paper, we discuss an increasingly discussed area among the researchers, real-time power system stability assessment. It is really important to find a reliable method which can work in real-time because of the increasing use of the power systems at the stability limit. We discussed the most traditional and trusted methods, the static method, these methods iteratively find the solution for the voltage stability nose point. But the problem with these methods is that they cannot be applied in real-time as they take a lot of time to be computed for a large number of buses. Therefore to make a reliable real-time power stability assessment system we discussed and compared the much faster and reliable methods, such as thevenin's Equivalent methods, and Artificial Neural Network methods.

\end{document}